\documentclass[showpacs,pre]{revtex4}
\usepackage{graphicx}
\begin{document}

\title{Two-dimensional dark soliton in the nonlinear Schr\"odinger equation}

\author{Hidetsugu  Sakaguchi and Tomoko Higashiuchi}

\address{Department of Applied Science for Electronics and Materials,\\ Interdisciplinary Graduate School of Engineering Sciences,\\
Kyushu University, Kasuga, Fukuoka 816-8580, Japan}

\begin{abstract}
Two-dimensional gray solitons to the nonlinear Schr\"odinger equation are numerically created by two processes to show its robustness. 
One is transverse instability of a one-dimensional gray soliton, and another is a pair annihilation of a vortex and an antivortex. 
The two-dimensional dark solitons are anisotropic and propagate in a certain direction. The two-dimensional dark soliton is stable against the head-on collision.  The effective mass of the two-dimensional dark soliton is evaluated from the motion in a harmonic potential.  
\end{abstract}
\maketitle
\section{Introduction}
The nonlinear Schr\"odinger equation is a typical soliton equation. Solitons in optical fibers are described by the nonlinear Schr\"odinger equation \cite{rf:1}. Recently, solitons, dark solitons and vortices were found in the Bose-Einstein condensates \cite{rf:2,rf:3,rf:4,rf:5,rf:6,rf:7}. The matter-wave solitons are also described with the nonlinear Schr\"odinger equation or the Gross-Pitaevskii equation. 
The one-dimensional nonlinear Schr\"odinger equation is an integrable system, and the time evolution can be solved exactly. The two-dimensional nonlinear 
Schr\"odinger equation has been intensively studied but the behaviors are not understood completely. We study mainly a two-dimensional dark soliton to the two-dimensional nonlinear Schr\"odinger equation. The model equation is written as 
\begin{equation}
i\frac{\partial \phi}{\partial t}=-\left (\frac{1}{2m}\right )\nabla^2\phi+g|\phi|^2\phi,
\end{equation}
where the Planck constant $\hbar$ is set to be 1, the coefficient $g$ of the nonlinear term can be set to  +1 or -1, and the mass parameter $m$ is assumed to be $m=1$ for the sake of simplicity. 
There is a two-dimensional soliton called the Townes soliton in the two dimensional nonlinear Schr\"odinger equation in the focusing case of $g=-1$, but the Townes soliton is unstable. The peak amplitude of the localized solution increases and it leads to divergence, if the norm of the localized solution is larger than the norm of the Townes soliton. Hereafter, we consider the defocusing case of $g=1$. There is a family of one-dimensional dark soliton to the two-dimensional nonlinear Schr\"odinger equation, which has a form
\begin{equation}
\phi(x,y)=\{iB+A\tanh A(y-vt)\}\exp(-i\mu t), 
\end{equation}
where $A^2+B^2=1$ and $v=B$. In the case of $B=0$, the minimum value of $|\phi|$ becomes zero and it is called a black soliton. For $B\ne 0$, the minimum value of $|\phi|$ is $|B|$ and then the dark soliton is called a gray soliton. The gray soliton propagates with velocity $B$. Because the nonlinear Schr\"odinger equation is invariant with respect to the scale transformation: $\phi\rightarrow \alpha \phi,\;x\rightarrow x/\alpha,\;y\rightarrow y/\alpha,\; t\rightarrow t/\alpha^2$, $\phi=\alpha\{iB+A\tanh A\alpha(y-v\alpha t)\}\exp(-i\mu \alpha^2 t)$ is also a dark soliton.  
The velocity of the dark soliton is scaled as $v\rightarrow\alpha v$. There is another family of two dimensional solutions called vortex, which has a form $\phi=U(r)\exp(i m\theta)$, where $r$ is the radius from the vortex center, $\theta$ is the phase angle around the vortex center, and $m$ is a nonzero integer. 
The vorticity of this vortex solution is $2\pi m$. The quantized vorticity is characteristic of the quantum fluid. The amplitude $U(r)$ satisfies
\begin{equation}
\frac{1}{2}\left (\frac{dU}{dr^2}+\frac{1}{r}\frac{dU}{dr}-\frac{m^2}{r^2}U\right )+(1-U^2)U=0
\end{equation}
The amplitude $U(r)$ at the vortex center is zero, i.e., $U(0)=0$, and $U(r)=1$ for $r\rightarrow \infty$.
The vortex with $m=\pm 1$ is called a fundamental vortex.
The vortex with amplitude $\alpha$ is obtained by the scale transformation as $
\phi=\alpha U(\alpha r)\exp(i m\theta)$.
A two-dimensional dark soliton, which has a two-dimensional hole around the center, might be interpreted as a kind of vortex solution corresponding to $m=0$. However, it is known that there is no two-dimensional isotropic black soliton of the form $\phi(x,y)=U(r)$, which has a property of $U(0)=0$ and $v=0$. 
Jones and Roberts found numerically two-dimensional special solutions called the rarefaction pulses, which correspond to two-dimensional gray solitons \cite{rf:8,rf:9}.  In this paper, we study a few natural creation processes of the two dimensional gray solitons. Then, we show the stability of the two-dimensional dark soliton against the head-on collision. Finally we evaluate the effective mass of the two-dimensional dark soliton from the motion in a harmonic potential.    

\section{Instability of a one-dimensional dark soliton and creation of a two-dimensional dark soliton}
The one-dimensional dark soliton is linearly unstable for the modulation in the $x$-direction. It is observed in numerical simulations and experiments that the one-dimensional dark soliton breaks up into vortex-antivortex pairs due to the development of the transverse instability \cite{rf:10,rf:11,rf:12}. We have performed numerical simulations again by changing the $B$ value of the initial dark soliton. The system size is set to be $L_x\times L_y=25\times 200$. We performed numerical simulations with the spit-step Fourier method of $256\times 2048$ modes. The periodic boundary conditions are imposed. The initial conditions are assumed to $\phi=iB-b\cos\{2\pi (x-L_x/2)/L_x\}{\rm sech}\{A(y-L_y/8)\}+A\tanh\{A(y-L_y/8)\}$ for $y<L_y/2$ and $\phi=iB-b\cos\{2\pi (x-L_x/2)/L_x\}{\rm sech}\{A(7L_y/8-y)\}+A\tanh\{A(7L_y/8-y)\}$ for $y>L_y/2$, where $A=\sin\theta_0$, $B=\cos\theta_0$, and $b=0.15$ is used.  That is, two one-dimensional dark solitons are initially set at $y=L_y/8$ and $7/8L_y$, and transverse modulations are overlapped on the one-dimensional dark soliton solution. The black soliton corresponds to the case of $\theta_0=\pi/2$. As $\theta_0$ is decreased, the minimum value of the dark soliton becomes larger and the propagation velocity increases. The initial condition is mirror-symmetric with respect to $y=L_y/2$ and $x=L_x/2$. For $\theta_0>2\pi\cdot 42/360$, the transverse modulation increases and a pair of vortex and antivortex are created. Figure 1(a) displays a time evolution of $|\phi(x,y_{min})|$ for $\theta_0=2\pi\cdot 45/360$. Here,  $y_{min}$ is the $y$ coordinate where $|\phi|$ takes the minimum value.   The transverse modulation grows, and two minimum points appear, which corresponds to the appearance of a vortex pair. Figure 1(b)  displays a time evolution of the $x$-coordinate of the vortex centers for $\theta_0=2\pi\cdot 45/360$.  The positions of the vortex centers were determined by the phase singularity points of $\phi(x,y)$ in our numerical simulations. The vortex pair is created at $x=L_x/2=12.5$, and the vortex pair is rapidly separated away.
    
\begin{figure}[tbp]
\includegraphics[height=4.cm]{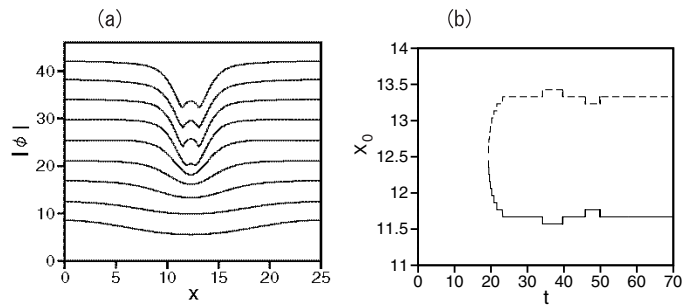}
\caption{(a) Time evolution of $|\phi(x,y_{min})|$ for $\theta_0=2\pi\cdot 45/360$. (b) Time evolution of vortex centers of the vortex and antivortex pair for $\theta_0=2\pi\cdot 45/360$.}
\label{fig1}
\end{figure}
\begin{figure}[tbp]
\includegraphics[height=6.cm]{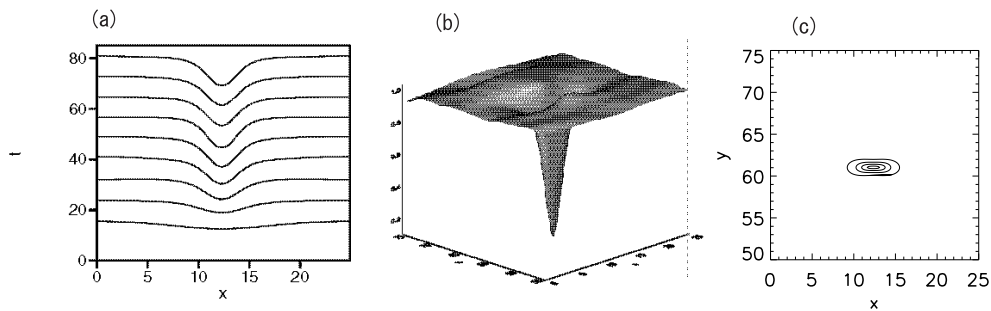}
\caption{Two dimensional dark soliton for $\theta_0=2\pi/9$. (a) Time evolution of $|\phi(x,y_{min})|$. (b) Three-dimensional plot of the two-dimensional dark soliton. (c) Contour plot of the two-dimensional dark soliton at $t=56$ starting from the initial condition with $\theta_0=2\pi/9$.}
\label{fig2}
\end{figure}
\begin{figure}[tbp]
\includegraphics[height=4.cm]{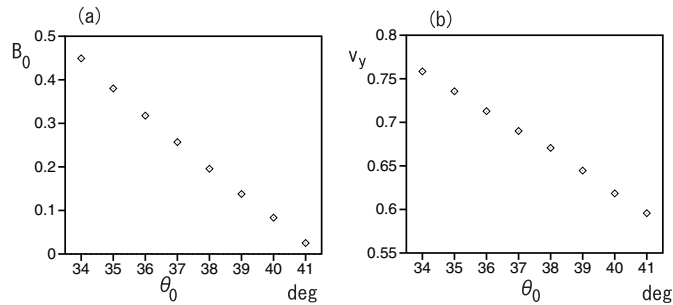}
\caption{Minimum value $B$ of $|\phi|$ as a function of $\theta_0$(deg.). (b) Velocity in the $y$ direction as a function of $\theta_0$(deg.).
}
\label{fig3}
\end{figure}
For $\theta_0<2\pi\cdot 41/360$, the growth of the transverse modulation saturates in time and a pair of vortex is not created as shown in Fig.~2(a).  The transverse modulation leads to the formation of a two-dimensional dark soliton. Figure 2(b) is a three dimensional plot of $|\phi|$ at $t=56$ for $\theta_0=2\pi\cdot 40/360$. The minimum value of $|\phi|$ in the two-dimensional dark soliton is about 0.08. 
We defines $B_0$ as the minimum value of $|\phi|$ in the two-dimensionl dark soliton, which is different from the minimum value $B$ of $|\phi|$ in the initial one-dimensional dark soliton.
The nozero value of $B_0$ implies a gray soliton. 
It is propagating in the $y$-direction. Figure 2(c) displays a contour map of $|\phi|$ at $t=56$. The two-dimensional dark soliton is anisotropic. The width in the $y$ direction is shorter than that in the $x$ direction. 

We have calculated the minimum value $B_0$ and the velocity $v_y$ in the $y$-direction as a function of the parameter $\theta_0$ in the initial condition. The results are shown in Fig.~3. The minimum value $B_0$ increases almost linearly from 0, as $\theta_0$ is decreased from the critical value of the formation of the two-dimensional dark soliton. The velocity $v_y$ also increases almost linearly as $\theta_0$ is decreased.  The velocity $v_y$ is approximately expressed as $v_y\sim 0.38\cdot B_0+0.59$.   This is a relation comparable to the relation $v_y=B$  for the one-dimensional dark soliton.   There is no two-dimensional dark soliton with velocity $v_y<0.59$.

\section{Merging of a pair of vortices and creation of a two-dimensional dark soliton}
The two-dimensional dark soliton might be also interpreted as a merged state of the vortex and antivortex. We have investigated time evolutions of vortex and antivortex pairs.   
The vortex pair interacts with each other. 
If the distance of the two vortices is sufficiently large, the vortices are expected to move as point vortices.
The equation of motion of the $i$th point vortex among $N$ vortices is generally described as \cite{rf:13}
\begin{equation}
\frac{dX_i}{dt}=\frac{\partial \Psi_i}{\partial Y_i},\frac{dY_i}{dt}=-\frac{\partial \Psi_i}{\partial X_i},\;\Psi_i=-\frac{1}{2\pi}\sum_{j\ne i}\Gamma_j\ln r_{i,j},
\end{equation}
where $X_i$ and $Y_i$ are the $x$ and $y$ coordinates of the vortex center of the $i$th vortex, $\Gamma_j$ is the vorticity of the $j$th vortex and $r_{ij}=\sqrt{(X_i-X_j)^2+(Y_i-Y_j)^2}$.  In the simplest case of a single vortex-antivortex pair with quantized vorticity $\Gamma_j=\pm 2\pi$, the equation of motion of the vortex centers obeys $dY_0/dt=1/(2x_0)$ and $dX_0/dt=0$, if a vortex and an antivortex are initially set at $(L_x/2-x_0,L_y/2)$ and $(L_x/2+x_0,L_y/2)$. The vortex pair propagates  in the $y$-direction, but does not move in the $x$-direction.  
However, if the separation of the two vortices is sufficiently small, the assumption of the point vortex is not good. The typical length scale is about $1/\alpha$ for the vortex solution $\phi=\alpha U(\alpha r)\exp(\pm i\theta)$. The assumption of the point vortex becomes worse, if the distance between the two vortices is smaller than $2/\alpha$. 
We have performed numerical simulations of the motion of a vortex-antivortex pair. A vortex and an antivortex with $\alpha=0.347$ are used as an initial condition. We have confirmed that the vortex pair propagates in the $y$ direction with  nearly $v_y=1/(2x_0)$ for large $x_0$. However, if the initial distance of the vortex pair is smaller than $x_{0c}=2.3$, the vortex pair approaches each other and it leads to the pair annihilation. We have also observed complicated behaviors such that  the vortex pair reappear in a certain time after the pair annihilation.  

\begin{figure}[tbp]
\includegraphics[height=6.cm]{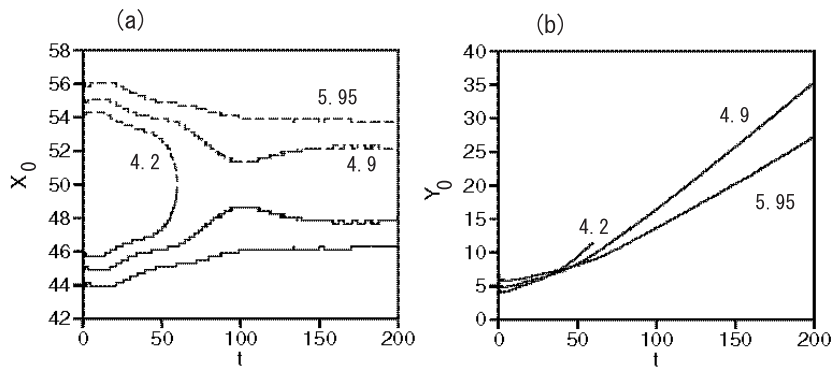}
\caption{Time evolutions of (a) $x-$coordinate $X_0$ and (b) $y-$coordinate $Y_0$ of vortex centers.}
\label{fig4}
\end{figure}
\begin{figure}[tbp]
\includegraphics[height=4.5cm]{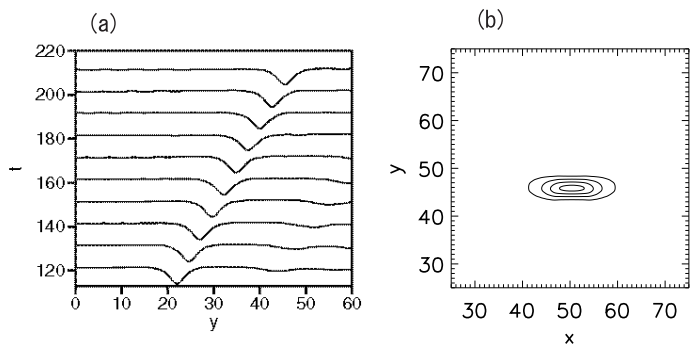}
\caption{(a) Time evolution of $|\phi(L_x/2,y)|$ for $x_0=4.2$. (b) Contour 
plot of $|\phi(x,y)|$ at $t=200$ for $x_0=4.2$.}
\label{fig5}
\end{figure}
We investigate the interaction among four vortices more in detail. 
If two vortices are set at $(-X_0,Y_0)$ and $(X_0,-Y_0)$ and two antivortices are set at $(X_0,Y_0)$ and $(-X_0,-Y_0)$, the four vortices interact with one another.  If the mirror symmetries with respect to $x=0$ and $y=0$ are satisfied during the time evolution, the equation of motion of the vortex center $(X_0,Y_0)$ of the vortex in $x>0$ and $y>0$ is expressed as
\begin{eqnarray}
\frac{dY_0}{dt}&=&\frac{1}{2X_0}-\frac{X_0}{2(X_0^2+Y_0^2)},\nonumber\\
\frac{dX_0}{dt}&=-&\frac{1}{2Y_0}+\frac{Y_0}{2(X_0^2+Y_0^2)},
\end{eqnarray} 
from the theory of point vortex \cite{rf:13}. 
The solution of the point vortices satisfies the relation $4X_0^2Y_0^2=a^2(X_0^2+Y_0^2)$, where $a$ is an arbitrary constant.  If the initial conditions are $X_0(0)=Y_0(0)=x_0$, $a$ is equal to $\sqrt{2}x_0$, and then $X_0(t)$ decreases to $x_0/\sqrt{2}$ from $X_0(0)=x_0$, and $Y_0(t)$ increases to infinity from $Y_0(0)=x_0$, as $t\rightarrow \infty$. The distance between the vortex and the antivortex approaches a smaller value $\sqrt{2}x_0$ owing to the interaction among the four vortices, and the final velocity of $Y_0$ takes a faster value of $v_y=1/(\sqrt{2}x_0)$, compared to a single vortex pair.   We have performed direct numerical simulations of the two pairs of vortex and antivortex. Four vortices are set at $(L_x/2-x_0,x_0)$, $(L_x/2+x_0,x_0)$, $(L_x/2-x_0,L_y-x_0)$, and $(L_x/2+x_0,L_y-x_0)$ as an initial condition. The two pairs of vortex and antivortex are located near the boundaries $y=0$ and $y=L_y$, and they are set to be mutually mirror-symmetric with respect to the lines $y=0$ and $y=L_y$ under the periodic boundary conditions.  The four vortices therefore interact with one another through the boundaries at $y=0$ and $y=L_y$. 
 The amplitude of the vortices is $\alpha=0.347$, and the system size is $(L_x\times L_y)=(100,200)$.
Figure 4(a) displays time evolutions of the $x$ coordinates of the vortex and antivortex in the region of $y<L_y/2$ for three initial values $x_0=4.2$, 4.9 and 5.95.   For $x_0=4.2$, the vortex-antivortex pair is annihilated at $t=60.1$.    The reappearance of the vortex pair was not observed. The critical value $x_{0c}=4.3$ of $x_0$ for the pair annihilation is larger than the critical value $x_{0c}=2.3$ in the case of a single vortex pair.  This is owing to the interaction among the four vortices.   For $x_0=4.9$, the vortex pair approaches once, but they separate away.  For $x_0=5.95$, the vortex pair approaches initially but the distance keeps about 7.5 for large $t$, which is close to the value $\sqrt{2}x_0\sim 8.4$ expected from the motion of the point vortex. Figure 4(b) displays time evolutions of the $y$ coordinates $Y_0$ of the vortex center for the same initial values  $x_0=4.2$, 4.9 and 5.95. For $x_0=5.95$, the vortex pair moves with a constant velocity $v_y\sim 0.135$, which is close to the velocity $1/7.5=0.133$ of a point vortex pair. 
As $x_0$ is decreased, the distance between the vortex pair is decreased, and therefore, the velocity in the $y$ direction is increased. 
For $x=4.9$, the average distance between the vortex pair for $150<t<200$ is 4.4, and the velocity $v_y$ is about 0.195, which is slightly smaller than $1/4.4=0.22$. The point vortex approximation might become worse, if the distance between the vortex pair is small. For $x_0=4.2$, the vortex pair is annihilated at $t\sim 60.1$, and the curve of $Y_0$ stops at the time.  

Figure 5(a) displays a time evolution of $|\phi(L_x/2,y)|$ for $x_0=4.2$.
The merging of the vortex pair leads to the formation of a two-dimensional dark soliton. The two-dimensional dark soliton propagates with a constant velocity $v_y\sim 0.26$. Figure 5(b) displays a contour plot of $|\phi(x,y)|$ at $t=200$ for $x_0=4.2$.  An anisotropic structure of the dark soliton 
is seen also in this case. The minimum value of $|\phi|$ of the two-dimensional dark soliton is nearly 0.14. The minimum value corresponds to $B_0=0.14\cdot 1/0.347=0.403$ for the vortex with $\alpha=1$. The velocity of the dark soliton with $\alpha=1$ and $B_0=0.403$ is estimated as $0.743$ from the relation $v_y\sim 0.38\cdot B_0+0.59$ obtained from Fig.~3. The corresponding velocity of the dark soliton for $\alpha=0.347$ is expected to be $v_y=0.743\alpha=0.743\cdot 0.347\sim 0.26$ from the scaling relation, which is nearly the same value as the velocity measured in the direct numerical simulation.  It implies that the two-dimensional dark soliton obtained by the annihilation of a vortex pair is the same as the two-dimensional dark soliton obtained by the transverse instability of a one-dimensional dark soliton.

\section{Collision of two-dimensional dark solitons}
We study the head-on collision of two two-dimensional dark solitons, to show the stability of the two-dimensional dark soliton. 
The system size is again $L_x\times L_y=25\times 200$.
As an initial condition, two one-dimensional dark solitons of $\theta_0=2\pi/9$ are set at $y=L_y/8$ and $7/8 L_y$, which are mirror-symmetric with respect to $y=L_y/2=100$. 
The time evolution of $\phi(L/2,y)$ of two dark solitons is shown in Fig.~6. 
The one-dimensional dark solitons evolve into two-dimensional dark solitons after a while. The two-dimensional dark soliton in $y<L_y/2=100$ propagates in the $+y$ direction and it collides with its mirror-symmetric two-dimensional dark soliton propagating in the $-y$ direction at $t\sim 120$. The two dark solitons pass through each other after the collision. It implies that the two-dimensional dark solitons are stable against the head-on collision. That is, the two-dimensional dark soliton has a property of "soliton". 

\begin{figure}[tbp]
\includegraphics[height=7cm]{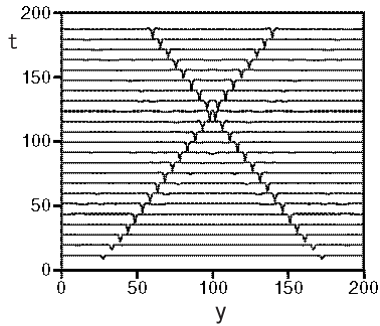}
\caption{Head-on collision of couter-propagating two-dimensional dark solitons starting from one-dimensional dark solitons with $\theta_0=2\pi/9$. Time evolution of $|\phi|$ at the cross section $x=L_x/2$ is shown.}
\label{fig6}
\end{figure}
\begin{figure}[tbp]
\includegraphics[height=8.cm]{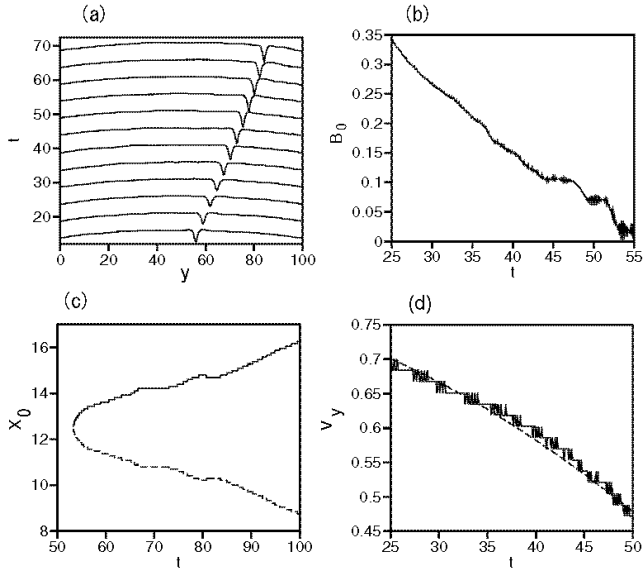}
\caption{(a) Time evolution of $|\phi(L_x/2,y)|$ of a two-dimensional dark soliton in a harmotic potential $U(y)=0.00025\cdot y^2$. (b) Time evolution of the minimum value $B_0$. (c) Time evolution of the $x$ coordinates $X_0$ of a vortex pair. (d) Velocity $v_y$ of the two-dimensional dark soliton in the $y$ direction. The dashed line is $v_y=0.78\cos(0.018t)$.}
\label{fig6}
\end{figure}
\section{Motion of a two-dimensional dark soliton in a harmonic potential}
We study a dynamical behavior of the two-dimensional dark soliton in a harmotic potential. 
It is known that the one-dimensional dark soliton behaves like a classical-mechanical particle in a weak harmonic potential $U(y)=ky^2/2$, and the center $y$ of the dark soliton obeys the equation 
\[m_{eff}\frac{d^2y}{dt^2}=-ky.\]
However, the effective mass $m_{eff}$ is not $m=1$, which is simply expected from  Eq.~(1), but nearly 2 \cite{rf:14,rf:15,rf:16}.  
On the other hand, the center of the bright soliton and the vortex obey the equation of motion with $m_{eff}=1$ in a weak harmonic potential \cite{rf:17}. The motion of the background  density needs to be taken into considered to understand the effective mass different from  1.  The difference of $m_{eff}$ and $m=1$ is also related to the fact that the $B$-value and therefore the velocity $v_y=B$ can take an arbitrary value in the dark soliton solution (2). The $B_0$-value in the two-dimensional dark soliton  also takes an arbitrary value, and therefore the velocity $v_y$ changes continuously according to the $B_0$-value as shown in Fig.~3.     
We can therefore expect that the effective mass of the two-dimensional dark soliton might be different from 1. To check the possibility, we have studied a model equation: 
\begin{equation}
i\frac{\partial \phi}{\partial t}=-\left (\frac{1}{2}\right )\nabla^2\phi+|\phi|^2\phi+U(y)\phi,
\end{equation}
where $U(y)=k/2(y-L_y/4)^2$ for $y<L_y/2$ and $U(y)=k/2(y-3L_y/4)^2$ for $y>L_y/2$ with $k=0.0005$. The potential also satisfies the mirror-symmetry with respect to $y=L_y/2=100$. As an initial condition, we have set a two-dimensional dark soliton at $y=L_y/4=50$, which was  obtained due to the transverse instability of a one-dimensional dark soliton of $\theta_0=2\pi/360\cdot 34$. 
Figure 7(a) displays a time evolution of $|\phi|$ at the cross section of $x=L_x/2$. Figure 7(b) displays the time evolution of the minimum value $B_0$ of $|\phi|$. The $B_0$-value decreases with $t$ owing to the harmonic potential.  When the $B_0$ value reaches 0, the two-dimensional dark soliton cannot exist and a pair of vortices is created from the two-dimensional dark soliton. Figure~7(c) displays a time evolution of the $x$ coordinate of the vortices.
The creation of the vortex pair occurs at $t\sim 53$. Figure 7(d) displays the time evolution of the velocity $v_y$ of the two-dimensional dark soliton in the $y$-direction before the creation of the vortex pair. The velocity $v_y$ is approximated as $v_y\sim 0.78\cos(0.0182t)$. 
For a classical-mechanical particle with mass $m_{eff}$ in the harmonic potential, the velocity obeys $v_y=v_0\cos\{(k/m_{eff})^{1/2}t\}$. Figure 7(d) implies that the two-dimensional dark soliton behaves like a classical-mechanical particle in the harmonic potential, however, the effective mass $m_{eff}$ is evaluated as $m_{eff}\sim k/0.0182^2\sim 1.5$.  Thus, we have confirmed that the effective mass of the two-dimensional dark soliton is different from 1, although we have not yet succeeded in calculating the effective mass theoretically.    

\section{Summary}
We have numerically created two-dimensional dark solitons using two kinds of processes. One is the transverse instability of one-dimensional dark solitons, and the other is the pair annihilation of a vortex and an antivortex.  The two-dimensional dark solitons are propagating gray solitons. The dark soliton is anisotropic in space , i.e. the width of the amplitude is smaller in the direction of the propagation.  We have shown that the two-dimensional dark soliton is stable against the collision.  Finally, we have investigated the dynamical behavior of the two-dimensional dark soliton in a weak harmonic potential and found that the effective mass is different from 1.

\end{document}